\documentclass[12pt,a4paper]{article}
\usepackage{amsmath}
\usepackage{amssymb}
\usepackage{amsthm}
\usepackage{latexsym}
\usepackage[english,russian]{babel}

\newcommand{\im}{\mathop{\rm Im}\nolimits}

\newcommand{\diag}{\mathop{\rm diag}\nolimits}
\newcommand{\sgn}{\mathop{\rm sgn}\nolimits}

\numberwithin{equation}{section} \swapnumbers

\newtheorem{Proposition}{Предложение}[section]

\theoremstyle{definition}
\newtheorem{defn}{Определение}[section]
\theoremstyle{remark}

\def\?{\marginpar{$\bullet\bullet\bullet$}}

\title{\bf О некоторых алгебраических конструкциях на произведении сфер}
\author{Игорь~Баяк}
\date{}

\begin{document}

\maketitle
\begin{abstract}

В данной работе мы исследуем топологическую связь между абелевыми
и неабелевыми группами четности. Абелевы группы четности
формируются как ядра гомоморфизмов четности в группе
$\mathbb{Z}^{n}$ а неабелевы группы четности формируются как ядра
гомоморфизмов четности в группе $S_{2}\wr S_{n}$. Факторизацией
целочисленной решетки с помощью абелевых групп четности мы
получаем фактор-решетки, которые формируют каркас (1-мерный
клеточный комплекс) произведения сфер. Показано, что изоморфизмы
этих фактор-решеток, образуют соответствующие неабелевы группы
четности.

\end{abstract}

\section{Группа $S_{2}\wr S_{n}$ и гомоморфизмы четности}

Хорошо известно, что группа подстановок $S_{n}$ допускает
расширение до группы $P_{n}=S_{2}\wr S_{n}$, которую легко
представить группой таких линейных преобразований
$\mathbb{R}^{n}\rightarrow\mathbb{R}^{n}:
(x_{1},\ldots,x_{n})\rightarrow(x'_{1},\ldots,x'_{n}): x'_{i}=\pm
x_{j}$, в которых отображение множества индексов координат
биективно, или группой квадратных матриц порядка $n$, имеющих в
каждом столбце и в каждой строке по одному ненулевому элементу
равному $1$ или $-1$.

Однако малоизвестно, что на группе $P_{n}$ можно задать три типа
функций четности. Действительно, по определению $S_{2}\wr
S_{n}=\prod^{n}S_{2} \leftthreetimes S_{n}$, где группа $S_{n}$
действует на группе $\prod^{n}S_{2}$ подстановками компонент
прямого произведения. Поэтому всякий элемент $z\in S_{2}\wr S_{n}$
раскладывается в произведение $z=xy$, где $x\in S_{n}$ и $y =
(y_{1},\ldots,y_{n})\in\prod^{n}S_{2}$, а сопряженный ему элемент
$z^{*}\in S_{2}\wr S_{n}$ раскладывается в произведение
$z^{*}=yx$. Тогда можно сделать следующие определения четности
элемента $z$ и сопряженного ему элемента $z^{*}$.
\begin{defn}
Функцией четности первого типа называется функция $\sgn z= \sgn
z^{*} = \sgn x$
\end{defn}
\begin{defn}
Функцией четности второго типа называется функция $\sgn z=\sgn
z^{*} = \sgn y = \sgn y_{1}*\cdots*\sgn y_{n}$
\end{defn}
\begin{defn}
Функцией четности третьего типа называется функция $\sgn z= \sgn
z^{*} = \sgn x*\sgn y$
\end{defn}
Все эти функции гомоморфно отображаются в группу $\{\pm1\}$.
Действительно, пусть дано разложение $x=x'x''$ и $y=y'y''$. Тогда,
если $\sgn xy=\sgn x$, то $\sgn x'y'*\sgn x''y''=\sgn x'*\sgn
x''=\sgn x'x''= \sgn x= \sgn xy$, если $\sgn xy=\sgn y$, то $\sgn
x'y'*\sgn x''y''=\sgn y'*\sgn y''=\sgn y'y''= \sgn y= \sgn xy$,
если $\sgn xy=\sgn x* \sgn y$, то $\sgn x'y'*\sgn x''y''=\sgn x'*
\sgn y'* \sgn x''* \sgn y''=\sgn x'* \sgn x''* \sgn y'* \sgn y''=
\sgn x* \sgn y = \sgn xy$, чем и доказывается, что все наши
функции четности являются гомоморфизмами. Таким образом,
гомоморфизмы четности выделяют в группе $P_{n}$ три подгруппы:
$AP_{n}$, $BP_{n}$, $CP_{n}$, которые формируются как ядра функций
четности соответственно первого, второго и третьего типа.

Исследуем теперь алгебраическую структуру этих групп. Прежде всего
напомним, что группа $P_{n}$ изоморфна группе
$\mathbb{Z}^{n}_{2}\leftthreetimes S_{n}$, где
$\mathbb{Z}^{n}_{2}$ это прямое произведение $n$ компонент
двухэлементного поля $\mathbb{Z}_{2}$, и заметим, что гомоморфизм
четности группы $\mathbb{Z}^{n}_{2}$, равный сумме по модулю 2
всех $n$ компонент ее элемента, выделяет в ней подгруппу
$A\mathbb{Z}^{n}_{2}$, состоящую из элементов, в которых 1
встречается четное число раз или вовсе не встречается. Поскольку
всякий элемент группы $A\mathbb{Z}^{n}_{2}$ раскладывается в
сумму, каждое слагаемое которой состоит из пар единиц и остальных
нулей, то группа $A\mathbb{Z}^{n}_{2}$ порождается своими
подгруппами, изоморфными $A\mathbb{Z}^{2}_{2}$. Тогда из
определения групп $AP_{n}$, $BP_{n}$ следует, что группа $AP_{n}$
изоморфна группе $\mathbb{Z}^{n}_{2}\leftthreetimes A_{n}$, где
$A_{n}$ это знакопеременная группа, а группа $BP_{n}$ изоморфна
группе $A\mathbb{Z}^{n}_{2}\leftthreetimes S_{n}$. Поскольку
знакопеременная группа порождается циклами длины 3, то группа
$AP_{n}$ порождается своими подгруппами третьей степени $AP_{3}$.
В свою очередь, поскольку $S_{n}$ порождается транспозициями а
группа $A\mathbb{Z}^{n}_{2}$ порождается своими двухкомпонентными
подгруппами, то группа $BP_{n}$ порождается своими подгруппами
второй степени $BP_{2}$. Выпишем здесь линейное представление и
матричный образ группы $BP_{2}$: $\{(x_{1},x_{2})\mapsto
(x_{1},x_{2}), (x_{1},x_{2})\mapsto
(x_{2},x_{1}),(x_{1},x_{2})\mapsto (-x_{1},-x_{2}),
(x_{1},x_{2})\mapsto (-x_{2},-x_{1})\}$,
$\left\{\left(\begin{array}{cc}
  1 & 0 \\
  0 & 1
\end{array}\right),\left(\begin{array}{cc}
  0 & 1 \\
  1 & 0
\end{array}\right),\left(\begin{array}{cc}
  -1 & 0 \\
  0 & -1
\end{array}\right),\left(\begin{array}{cc}
  0 & -1 \\
  -1 & 0
\end{array}\right)\right\}$.
Элементам этой группы соответствуют отражения плоскости
$(x_{1},x_{2})$ относительно диагоналей $x_{2}=x_{1}$ и
$x_{2}=-x_{1}$. Выпишем также линейное представление и матричный
образ группы $CP_{2}$: $\{(x_{1},x_{2})\mapsto (x_{1},x_{2}),
(x_{1},x_{2})\mapsto (x_{2},-x_{1}),(x_{1},x_{2})\mapsto
(-x_{1},-x_{2}), (x_{1},x_{2})\mapsto (-x_{2},x_{1})\}$,
$\left\{\left(\begin{array}{cc}
  1 & 0 \\
  0 & 1
\end{array}\right),\left(\begin{array}{cc}
  0 & 1 \\
  -1 & 0
\end{array}\right),\left(\begin{array}{cc}
  -1 & 0 \\
  0 & -1
\end{array}\right),\left(\begin{array}{cc}
  0 & -1 \\
  1 & 0
\end{array}\right)\right\}$.
Элементам этой группы соответствуют повороты евклидовой плоскости
на угол кратный $\pi/2$. Группа $CP_{n}$ также порождается своими
подгруппами, изоморфными $CP_{2}$. Действительно, возьмем
произвольный элемент группы $CP_{n}$ и, умножая его на элементы
группы, преобразующие только пару координат пространства
представления, получим сначала тождественное отображение в
компоненте $S_{n}$ а затем и в компоненте $S_{2}$ сплетенного
произведения, что возможно в силу свойства групп $S_{n}$ и
$A\mathbb{Z}^{n}_{2}$. Тогда, произведение обратных элементов
разложения будет равно исходному элементу $CP_{n}$, а
следовательно всякий элемент группы $CP_{n}$ можно разложить в
произведение элементов групп, изоморфных $CP_{2}$.

Пусть теперь дано такое разбиение множества $I=\{1,\ldots ,n\}$ на
непересекающиеся подмножества $I_{1},\ldots, I_{m}$, что мощности
подмножеств $I_{i}$ равны $n_{i}$, причем $n_{1}+\cdots +
n_{m}=n$. Тогда всякому разбиению $J=\{I_{1},\ldots,I_{m}\}$ можно
сопоставить группу, образованную внешним полупрямым произведением
\begin{equation}\label{JP}
  JP_{n}=CP_{n_{1}}\times\cdots\times CP_{n_{m}}\leftthreetimes BP_{m},
\end{equation}
где группа $BP_{m}$ действует на группе $\prod^{m}CP_{n_{i}}$
соответствующими подстановками компонент прямого произведения.
Назовем эту группу {\it конечной неабелевой группой четности}.
Следует отметить, что в общем случае группа $JP_{n}$ не имеет
матричного представления, но ее элементы могут быть представлены
матрицами (представляющими группу $BP_{m}$), ненулевые элементы
которых являются группами $CP_{n_{i}}$, также имеющими матричное
представление.

Пусть теперь задан матричный образ вещественного линейного
представления групп второй степени $CP_{2}$, $BP_{2}$. Обозначим
посредством $C$ и $B$ матричные алгебры, порождаемые матрицами,
соответствующих групп, а посредством $C^{*}$, $B^{*}$ обозначим
группы обратимых элементов соответствующих алгебр. Тогда несложно
показать, что ядро гомоморфизма четности $C^{*}\rightarrow
\mathbb{R}^{*}:C^{*}\rightarrow \det C^{*}$ равно специальной
ортогональной группе $SO(2)$, а ядро гомоморфизма четности
$B^{*}\rightarrow \mathbb{R}^{*}:B^{*}\rightarrow \det B^{*}$
равно специальной псевдоортогональной группе $SO(1,1)$.
Действительно, поскольку
  $C=\left\{\left(\begin{array}{cc}
  x & y \\
  -y & x
\end{array}\right)\right\}$, где $x,y\in\mathbb{R}$, то множество
решений уравнения $\det C^{*}=x^{2}+y^{2}=1$ как раз и выделяет в
группе $C^{*}$ подгруппу $SO(2)$. В свою очередь, поскольку
  $B=\left\{\left(\begin{array}{cc}
  x & y \\
  y & x
\end{array}\right)\right\}$, где $x,y\in\mathbb{R}$, то множество
решений уравнения $\det B^{*}=x^{2}-y^{2}=1$ выделяет в группе
$B^{*}$ подгруппу $SO(1,1)$. С помощью блочно-диагональных матриц
порядка $n$ мы зададим изоморфную $SO(2)$ группу $SO_{jk}(2)=\diag
\left[1,\ldots,SO(2)_{(jk)},\ldots,1\right]_{n}$, которая
образована матрицами, отличающимися от единичной только тем, что
на пересечении пары строк и пары столбцов с индексами $j$, $k$
находится матричный элемент группы $SO(2)$. Аналогично зададим
изоморфную $SO(1,1)$ группу $SO_{jk}(1,1)=\diag
\left[1,\ldots,SO(1,1)_{(jk)},\ldots,1\right]_{n}$. Тогда всякому
разбиению $J$ можно сопоставить группу
\begin{equation}\label{SO}
  SO(n_{1},\ldots,n_{m})=\left\langle SO_{jk}(2), SO_{jk}(1,1)
  \right\rangle_{J},
\end{equation}
порождаемую генераторами $SO_{jk}(2)$ и $SO_{jk}(1,1)$ так, что
пара индексов первого генератора принадлежит произвольному
подмножеству $I_{i}$, а пара индексов второго генератора
принадлежит произвольной паре подмножеств разбиения. Однако
заметим, что достаточное для образования группы
$SO(n_{1},\ldots,n_{m})$ число генераторов задается формулой
\begin{equation}\label{gen}
  p=\sum_{m}n_{i}(n_{i}-1)/2 + m(m-1)/2,
\end{equation}
поскольку для гиперболического поворота между двумя
подпространствами достаточно только одного генератора
$SO_{jk}(1,1)$ с произвольными индексами $j,k$, взятыми из двух
разных подмножеств $I_{i}$, а остальные генераторы $SO_{jk}(1,1)$
(с другими индексами) будут производными от выбранного генератора
и дискретных вращений внутри каждого из двух подпространств в
отдельности. Тем самым, поскольку группа Ли порождается своими
однопараметрическими подгруппами, то мы получили
$p$-параметрическую группу Ли $SO(n_{1},\ldots,n_{m})$, которую
назовем {\it группой четности Ли}.

Итак, взяв за основу группу $S_{2}\wr S_{n}$ и задав на ней
гомоморфизмы четности, нам удалось получить не только конечные но
и непрерывные группы четности, включающие в себя в том числе и
некоторые классические группы. Заметим однако, что для образования
всех возможных матричных групп Ли посредством порождения их
маломерными подгруппами Ли необходимо воспользоваться еще одной
матричной алгеброй, а именно, алгеброй
$A=\left\{\left(\begin{array}{cc}
  x & 0 \\
  0 & y
\end{array}\right)\right\}$, где $x,y\in\mathbb{R}$. Тогда, всякая
подгруппа общей линейной группы $GL(n,\mathbb{R})$ порождается
всевозможными маломерными группами Ли, которые могут быть
образованы из матричных алгебр $M(2,\mathbb{R})$, $A$,$B$,$C$ и
расширены группой мономиальных подстановок двухэлементного базиса.
Например, многосвязная, а точнее $2^{n}n!$-компонентная, группа
Ли, состоящая из $n$-матриц, в каждой строке и каждом столбце
которых по одному ненулевому элементу, порождается генераторами,
изоморфными односвязной группе Ли
$A^{+}=\left\{\left(\begin{array}{cc}
  x & 0 \\
  0 & y
\end{array}\right)\right\}$, где $x,y\in\mathbb{R}^{+}$, и
генераторами, изоморфными конечной группе $S_{2}\wr S_{2}$.

\section{Группа $\mathbb{Z}^{n}$ и гомоморфизмы четности}

Зададим на группе $\mathbb{Z}^{n}$ функцию четности
$\mathbb{Z}^{n}\rightarrow \mathbb{Z}_{2}: |x_{1}+\ldots+x_{n}|
\mod 2$, значение которой равно сумме по модулю 2 всех $n$
компонентов ее элемента. Тем самым, мы получим гомоморфизм группы
$\mathbb{Z}^{n}$ в группу $\mathbb{Z}_{2}$. Ядро гомоморфизма
четности мы обозначим $A\mathbb{Z}^{n}$ и назовем группой четных
элементов $\mathbb{Z}^{n}$. Легко заметить, что группа
$A\mathbb{Z}^{n}$ состоит из элементов $\mathbb{Z}^{n}$, в которых
нечетные компоненты встречаются четное число раз либо вовсе не
встречаются, и поэтому она порождается всевозможными своими
подгруппами второй степени $A\mathbb{Z}^{2}$. Имея ввиду, что
$A\mathbb{Z}^{1}= 2\mathbb{Z}$, сформируем также в
$\mathbb{Z}^{n}$ подгруппу $B\mathbb{Z}^{n}=(2\mathbb{Z})^{n}$,
состоящую из прямого произведения (суммы) $n$ экземпляров
$A\mathbb{Z}^{1}$, т. е. из четных целых во всех компонентах
$\mathbb{Z}^{n}$. Заметим при этом, что
$\mathbb{Z}^{n}/B\mathbb{Z}^{n}=\mathbb{Z}^{n}_{2}$.

Пусть далее мы имеем разбиение $J=\{I_{1},\ldots,I_{m}\}$,
определенное ранее. Тогда, в соответствии с этим разбиением можно
сформировать группу
\begin{equation}\label{JZ}
  J\mathbb{Z}^{n}=A\mathbb{Z}^{n_{1}}\times\cdots\times
  A\mathbb{Z}^{n_{m}}.
\end{equation}
Фактор-группу $\mathbb{Z}^{n}/J\mathbb{Z}^{n}$, изоморфную группе
$\mathbb{Z}_{2}^{m}$ порядка $2^{m}$ мы назовем {\it конечной
абелевой группой четности}. Порядок группы
$\mathbb{Z}^{n}/J\mathbb{Z}^{n}$ можно вычислить также, исходя из
того, что
\begin{equation}\label{Z/JZ}
  \mathbb{Z}^{n}/J\mathbb{Z}^{n}= \mathbb{Z}^{n_{1}}/A\mathbb{Z}^{n_{1}}
  \times\cdots\times  \mathbb{Z}^{n_{m}}/A\mathbb{Z}^{n_{m}}
\end{equation}
а всякая группа $\mathbb{Z}^{n_{1}}/A\mathbb{Z}^{n_{i}}$ состоит
из двух элементов. В качестве иллюстрации приведем здесь несколько
примеров. Если $J=\{(1),(2)\}$, то $J\mathbb{Z}^{2}=
\{2\mathbb{Z},2\mathbb{Z}\}$, а группа
$\mathbb{Z}^{n}/J\mathbb{Z}^{2}$ состоит из четырех точек. Если
$J=\{(1,2)\}$, то $J\mathbb{Z}^{2}= \{(2\mathbb{Z},2\mathbb{Z}),
(2\mathbb{Z}+1,2\mathbb{Z}+1)\}$, а группа
$\mathbb{Z}^{n}/J\mathbb{Z}^{2}$ состоит из двух точек, причем на
плоскости ее можно представить в качестве двух классов
целочисленных параллелограммов, т.е. параллелограммов,
составленных из пар целых чисел на его сторонах, с равноудаленными
от нулевой точки вершинами, лежащими в четных и нечетных точках
координатных осей соответственно.

Перейдем теперь к рассмотрению непрерывных абелевых групп. Пусть
даны гомоморфизмы $\mathbb{R}\rightarrow\mathbb{R}_{1}: (x\mapsto
|x|\mod 1)$ и $\mathbb{R}\rightarrow\mathbb{R}_{2}: (x\mapsto
|x|\mod 2)$, где $|x|\mod 1$ и $|x|\mod 2$ означают классы
сравнений числа $x$ по модулю 1 и 2 соответственно. Заметим при
этом, что если фактормножество
$\mathbb{R}_{2}=\mathbb{R}/2\mathbb{Z}$ изоморфно окружности
$S^{1}$, то фактормножество $\mathbb{R}_{1}=\mathbb{R}/\mathbb{Z}$
изоморфно проективной прямой $\mathbb{R}P^{1}$, которая получается
отождествлением противоположных точек окружности $S^{1}$, причем,
топологически $S^{1}$ и $\mathbb{R}P^{1}$ эквивалентны. Кроме
того, легко задать гомоморфизм $f:\mathbb{R}^{2}\rightarrow
\mathbb{R}_{2}\times\mathbb{R}_{2}: (x_{1},x_{2})\mapsto
(|x_{1}|\mod 2,|x_{2}|\mod 2)$, ядро которого формирует группу
$2\mathbb{Z}\times 2\mathbb{Z}$ а образ изоморфен тору
$S^{1}\times S^{1}$. Вместе с тем, можно задать гомоморфизм
$f:\mathbb{R}^{2}\rightarrow \mathbb{R}_{1}\times\mathbb{R}_{2}:
(x_{1},x_{2})\mapsto (|x_{2}|\mod 1,|x_{1}+x_{2}|\mod 2)$, ядро
которого формирует группу $A\mathbb{Z}^{2}$ а образ изоморфен
пространству сферических координат $\mathbb{R}^{2}(S^{2})$.
Действительно, достаточно установить соответствие между
сферическими координатами $(\phi, \theta)$ (где принято, что
широта $\phi$ измеряется по модулю $2\pi$, а долгота $\theta$ ---
по модулю $\pi$) и компонентами произведения $\mathbb{R}_{1}
\times \mathbb{R}_{2}$ согласно формул $\phi=\pi |x_{1}+x_{2}|\mod
2$, $\theta=\pi |x_{2}|\mod 1$, откуда сразу получим $\im f =
\mathbb{R}^{2}/ A\mathbb{Z}^{2}\simeq \mathbb{R}^{2}(S^{2})$.

Аналогично, можно задать гомоморфизм $f:\mathbb{R}^{n}\rightarrow
\mathbb{R}^{n-1}_{1} \times\mathbb{R}_{2}: (x_{1},\ldots,
x_{n})\mapsto (|x_{2}|\mod 1,\ldots, |x_{n}|\mod 1,
|x_{1}+\ldots+x_{n}|\mod 2)$, который имеет ядро, образующее
группу $A\mathbb{Z}^{n}$, а его образ изоморфен пространству
сферических координат $n$-мерной сферы $\mathbb{R}^{n}(S^{n})$,
т.е. $\im f = \mathbb{R}^{n}/ A\mathbb{Z}^{n}\simeq
\mathbb{R}^{n}(S^{n})$. Обобщение этой конструкции сводится к
тому, что для произвольного разбиения $J$ строится гомоморфизм,
имеющий в качестве образа {\it непрерывную абелеву группу
четности} $\mathbb{R}^{n}/J\mathbb{Z}^{n}$. А поскольку
\begin{equation}\label{R/JZ}
  \mathbb{R}^{n}/J\mathbb{Z}^{n}= \mathbb{R}^{n_{1}}/A\mathbb{Z}^{n_{1}}
  \times\cdots\times  \mathbb{R}^{n_{m}}/A\mathbb{Z}^{n_{m}},
\end{equation}
то непрерывная абелева группа четности изоморфна координатному
пространству прямого произведения сфер $S^{n_{1}}\times\cdots
\times S^{n_{m}}$.

\section{Соотношение между группами четности}

Прежде всего найдем дискретные автоморфизмы поля действительных
чисел $\mathbb{R}$, которые допускаются его факторизацией
$\mathbb{R}/2\mathbb{Z}$ и $\mathbb{R}/\mathbb{Z}$. Если мы
возьмем полярное представление комплексного числа $\rho
e^{i\varphi}$, где $\rho\in\mathbb{R}^{+}$, $\varphi\in[0,2\pi[$,
то его аргументу и модулю можно поставить в соответствие некоторые
линейные преобразования окружности $S^{1}=\mathbb{R}/2\mathbb{Z}$.
Так, модуль комплексного числа можно связать с коэффициентом
деформации окружности, т.е. модуль $\rho$ задает преобразование
$\mathbb{R}/2\mathbb{Z} \mapsto \rho\mathbb{R}/2\mathbb{Z}$, а его
аргумент можно связать с поворотами окружности, т.е. аргумент
$\varphi$ задает преобразование $\mathbb{R}/2\mathbb{Z}\mapsto
\mathbb{R}/2\mathbb{Z}+\varphi/\pi$. В свою очередь, если мы
возьмем полуполярное представление комплексного числа $\kappa
e^{i\theta}$, где $\kappa\in\mathbb{R}$, $\theta\in[0,\pi[$, то
его модуль $\kappa$ и аргумент $\theta$ задают соответствующие
линейные преобразования проективной прямой $\mathbb{R}P^{1}=
\mathbb{R}/\mathbb{Z}$, а именно: $\mathbb{R}/\mathbb{Z} \mapsto
\kappa\mathbb{R}/\mathbb{Z}$, $\mathbb{R}/\mathbb{Z}\mapsto
\mathbb{R}/\mathbb{Z}+\theta/\pi$. Таким образом, мы нашли искомые
дискретные автоморфизмы $\mathbb{R}/2\mathbb{Z} \mapsto
1\ast\mathbb{R}/2\mathbb{Z}$ и $\mathbb{R}/\mathbb{Z} \mapsto
\pm1\ast\mathbb{R}/\mathbb{Z}$, которые соответствуют дискретным
вращениям окружности и проективной прямой соответственно.

Пусть в пространстве $\mathbb{R}^{n}$ задана целочисленная решетка
$\mathbb{L}^{n}=\{\mathbb{Z}^{n};(\mathbb{Z},
\ldots,\mathbb{R},\ldots, \mathbb{Z})\}$, состоящая из прямых
линий, пересекающихся в точках $\mathbb{R}^{n}$ с целочисленными
координатами и параллельных базисным ортам. Целочисленная решетка
$\mathbb{L}^{n}$ представляет собой регулярно повторяющийся набор
узлов и ребер, который может быть получен бесконечным повторением
$n$-мерного образующего параллелепипеда, состоящего из $2^{n}$
узлов соединенных ребрами. Следовательно, группа изоморфизмов
целочисленной решетки совпадает с группой изоморфизмов образующего
параллелепипеда, и поэтому равна группе мономиальных подстановок,
которая изоморфна сплетенному произведению $S_{2}\wr S_{n}$.

Факторизуем теперь решетку $\mathbb{L}^{n}$ так, чтобы узлы
решетки факторизовались с помощью канонического гомоморфизма
$\mathbb{Z}^{n}\rightarrow \mathbb{Z}^{n}/\mathbb{Z}^{n}$, т.е. в
точку, а каждый класс параллельных прямых целочисленной решетки
факторизовался бы с помощью канонического гомоморфизма
$\mathbb{R}\rightarrow\mathbb{R}/\mathbb{Z}$ в проективную прямую
$\mathbb{R}P^{1}$. Потребуем при этом, чтобы касательные вектора к
проективным прямым факторизованной решетки в ее факторизованном
узле образовали систему из $n$ линейно независимых векторов. В
результате такой факторизации целочисленной решетки, мы получим
фактор-решетку $\mathbb{L}^{n}/ \mathbb{Z}^{n}$, у которой группа
изоморфизмов, сохраняющих неподвижной нулевую точку, равна
$S_{2}\wr S_{n}$. Это следует из того, что группа дискретных
вращений проективной прямой равна $S_{2}$ и нет никаких
топологических препятствий для образующих группу подстановок
$S_{n}$ перестановок 1-мерных элементов решетки
$\mathbb{L}^{n}/\mathbb{Z}^{n}$.

Установим теперь группу дискретных вращений фактор-решетки
$\mathbb{L}^{n}/A\mathbb{Z}^{n}$, т.е. группу изоморфизмов
фактор-решетки, сохраняющих неподвижной ее нулевую точку. Прежде
всего заметим, что фактор-решетка $\mathbb{L}^{1}/ 2\mathbb{Z}$
получается факторизацией регулярной решетки $\mathbb{L}^{1}$ в
результате канонического гомоморфизма $\mathbb{R}\rightarrow
\mathbb{R}/2\mathbb{Z}$ в окружность $S^{1}$, имеющую две
противоположные узловые точки. А поскольку дискретные вращения
окружности сводятся к ее тождественному отображению, то группа
дискретных вращений фактор-решетки $\mathbb{L}^{1}/2\mathbb{Z}$
тривиальна и равна $CP_{1}$. Если мы теперь образуем
фактор-решетку $\mathbb{L}^{2}/A\mathbb{Z}^{2}$, изоморфную пучку
из двух окружностей, пересекающихся в двух своих противоположных
точках, то легко установим, что группа дискретных вращений этого
пучка равна группе $CP_{2}$. Наконец, пусть дана фактор-решетка
$\mathbb{L}^{n}/A\mathbb{Z}^{n}$, изоморфная пучку из $n$
окружностей, пересекающихся в двух своих противоположных точках,
причем касательные вектора к окружностям в этих точках образуют
систему из $n$ линейно независимых векторов. Поскольку всякая пара
ее 1-мерных элементов равна $\mathbb{L}^{2}/ A\mathbb{Z}^{2}$, то
всякое дискретное вращение фактор-решетки $\mathbb{L}^{n}/
A\mathbb{Z}^{n}$ можно разложить в композицию дискретных вращений
всевозможных ее пар, а следовательно группа дискретных вращений
фактор-решетки $\mathbb{L}^{n}/ A\mathbb{Z}^{n}$ равна группе
четности $CP_{n}$.

Пусть дана фактор-решетка $\mathbb{L}^{2}/ B\mathbb{Z}^{2}$,
которую можно представить в виде четырех окружностей, точки
пересечения которых расположены в узлах образующего параллелограма
решетки $\mathbb{L}^{2}$. Поскольку всякая окружность этой решетки
не допускает обращения связанных ей узлов, то дискретные вращения
фактор-решетки $\mathbb{L}^{2}/ B\mathbb{Z}^{2}$ сводятся к тем
изоморфизмам образующего параллелограма, которые порождаются
исключительно его диагональными отражениями, а следовательно они
составляют группу $BP_{2}$. Аналогично, пусть дана фактор-решетка
$\mathbb{L}^{n}/ B\mathbb{Z}^{n}$, которую можно представить в
виде n-параллелепипеда, составленного из $2^{n}$ узлов, связанных
окружностями. Тогда, дискретные вращения фактор-решетки
$\mathbb{L}^{2}/ B\mathbb{Z}^{2}$ сводятся к тем изоморфизмам
образующего n-параллелепипеда, которые порождаются исключительно
его диагональными отражениями, а следовательно они составляют
группу $BP_{n}$.

Наконец, пусть дана фактор-решетка $\mathbb{L}^{n}/
J\mathbb{Z}^{n}$, которую можно представить в виде
m-параллелепипеда, составленного из $2^{m}$ узлов, связанных
пучками из $n_{i}$ окружностей. Поскольку дискретные вращения
фактор-решетки $\mathbb{L}^{n}/ J\mathbb{Z}^{n}$ сводятся к
составляющим группу $BP_{m}$ диагональным изомофизмам
m-параллелепипеда, которые действуют на прямом произведении
дискретных вращений пучков окружностей, составляющих группу
$\prod^{m}CP_{n_{i}}$, то ее группа дискретных вращений равна
$\prod^{m}CP_{n_{i}}\leftthreetimes BP_{m}$. Таким образом,
доказано
\begin{Proposition}
Группа дискретных вращений решетки
$\mathbb{L}^{n}/J\mathbb{Z}^{n}$ эквивалентна группе четности
$JP_{n}$
\end{Proposition}

Если рассмотреть целый континуум непрерывных вращений
фактор-решетки $\mathbb{L}^{n}/A\mathbb{Z}^{n}$, оставляющих на
месте ее узловые точки, то мы получим группу движений сферы
$S^{n}$, оставляющих неподвижными ее полюса, т.е., группу $SO(n)$.
Аналогично можно было бы получить группу движений прямого
произведения сфер, оставляющих неподвижными узловые точки
соответствующей фактор-решетки. Действительно, если фактор-решетку
$\mathbb{L}^{n} /J\mathbb{Z}^{n}$ рассматривать как каркас
(одномерный клеточный комплекс) прямого произведения сфер
$S^{n_{1}}\times\cdots \times S^{n_{m}}$, то группу дискретных
вращений фактор-решетки $JP_{n}$ можно было бы расширить до группы
четности Ли $SO(n_{1},\ldots,n_{m})$ и сопоставить ее группе
движений прямого произведения сфер $S^{n_{1}}\times\cdots \times
S^{n_{m}}$, оставляющих неподвижными узловые точки соответстующей
фактор-решетки.

В заключение воспользуемся полуполярным представлением комплексных
чисел, чтобы установить геометрическое представление компактной
унитарной группы $U(n)$. Пусть дано матричное представление группы
$U(n)$ в виде матриц $\left(\kappa_{ij} e^{\theta_{ij}}
\right)_{n}$, где $\kappa_{ij}\in \mathbb{R}$,
$\theta_{ij}\in[0,\pi[$. Тогда, приравнивая матричные компоненты
$\theta_{ij}$ нулевой матрице, мы получим из матричного
представления унитарной группы множество матриц
$(\kappa_{ij})_{n}$, составляющих ортогональную группу $O(n)$. С
другой стороны, приравнивая компоненты $\kappa_{ij}$ единичной
матрице, мы получим множество матриц $\diag [\theta_{ii}]_{n}$,
представляющих собой абелеву группу $T^{n}=\mathbb{R}^{n}
/\mathbb{Z}^{n}$. Вместе с тем, если взять такие произвольные
унитарные матрицы $A,B,C$, что $A$ и $C$ принадлежат подгруппе,
изоморфной $O(n)$, а матрица $B$ принадлежит подгруппе, изоморфной
$T^{n}$, то множество унитарных матриц $ABC$ составит группу,
имеющую размерность $n^{2}$, т.е. мы получим всю группу $U(n)$.
Следовательно имеет место
\begin{Proposition}
$U(n)\cong O(n) \cdot T^{n} \cdot O(n)$, где точка означает
связанное произведение.
\end{Proposition}
Заметим при этом, что двухлистным накрытием сферы $S^{2}$ с
выколотыми полюсами является тор $S^{1}\times S^{1}\cong
\mathbb{R}/2\mathbb{Z}\times\mathbb{R}/2\mathbb{Z}$, который
получается вращением окружности вокруг оси, касающейся в пределе
окружности в двух ее противоположных точках, т.е. так, что
окружность охватывает ось и в пределе стремится ее пересечь в двух
своих диаметрально противоположных точках. Отождествляя
противоположные точки этого тора, мы получаем группу
$T^{2}=\mathbb{R}^{2}/\mathbb{Z}^{2}$, а следовательно
геометрическое представление унитарной группы $U(2)$ сводится к
связанному произведению этой группы и групп вращения листов
накрытия сферы $S^{2}$ с выколотыми полюсами.

\end{document}